\documentclass{aastex}
\usepackage{emulateapj5,epsf,apjfonts}

\slugcomment{to appear in ApJ}

\shorttitle{The {\sc HELLAS2XMM} survey I.}
\shortauthors{Baldi et al.}

\begin{document}

\title{The {\sc HELLAS2XMM} survey: I. The X-ray data and the Log(N)-Log(S)}


\author{A. Baldi\altaffilmark{1}, S. Molendi}
\affil{Istituto di Fisica Cosmica - CNR, via Bassini 15, I-20133 Milano, Italy}
\email{baldi,silvano@ifctr.mi.cnr.it}

\author{A. Comastri}
\affil{Osservatorio Astronomico di Bologna, via Ranzani 1, I-40127 Bologna, 
Italy}
\email{comastri@bo.astro.it}

\author{F. Fiore}
\affil{Osservatorio Astronomico di Roma, via Frascati 33, I-00040 
Monteporzio, Italy}
\email{fiore@quasar.mporzio.astro.it}

\author{G. Matt}
\affil{Dipartimento di Fisica - Universit\'a di Roma Tre, via della Vasca 
Navale 84, I-00146 Roma, Italy}
\email{matt@fis.uniroma3.it}

\and

\author{C. Vignali\altaffilmark{2}}
\affil{Department of Astronomy and Astrophysics - The Pennsylvania
State University, 525 Davey Lab, University Park, PA 16802 USA}
\email{chris@astro.psu.edu}

\altaffiltext{1}{Dipartimento di Fisica - Universit\`a di Milano Bicocca, P.za della 
Scienza 3, I-20133 Milano, Italy}
\altaffiltext{2}{Dipartimento di Astronomia - Universit\'a di Bologna, via Ranzani 1, 
I-40127 Bologna, Italy}

\begin{abstract}
We present the first results from an {\em XMM-Newton} serendipitous medium-deep
survey, which covers nearly three square degrees.
We detect a total of 1022, 495 and 100 sources, down to minimum fluxes
of about $5.9\times10^{-16}$, $2.8\times10^{-15}$ and 
$6.2\times10^{-15}$ erg cm$^{-2}$ s$^{-1}$, in the 
0.5-2, 2-10 and 4.5-10 keV band, respectively. In the soft band this is one 
of the largest samples available to date and surely the largest in the 
2-10 keV band at our limiting X-ray flux.
The measured Log(N)-Log(S) are 
found to be in good agreement with previous determinations.
In the 0.5-2 keV band we detect a break at fluxes
around $5\times10^{-15}$ erg cm$^{-2}$ s$^{-1}$. In the harder bands, 
we fill in the gap at intermediate fluxes between
deeper {\em Chandra} and {\em XMM-Newton} observations and shallower 
{\em BeppoSAX} and {\em ASCA} surveys.
\end{abstract}



\section{Introduction}

In the last decade it has become progressively
clearer that the extragalactic X-ray 
background (XRB) originates from the superposition of many unresolved faint
sources.\\
In the soft band (0.5-2 keV) {\em ROSAT}
has resolved about 70\%-80\% of the XRB \citep{has98}, meanwhile recent 
{\em Chandra} deep
observations are resolving almost all the background 
\citep{mus00,gia01}. The hard band (2-10 keV) XRB has been resolved
at a 25\%-30\% level with {\em BeppoSAX} and {\em ASCA} surveys 
\citep*{cag98,ued99,gio00} and recently at more than 60\% with {\em Chandra} 
\citep{mus00,gia01,hor01}. Moreover, in the very hard band 
(5-10 keV) the fraction resolved
by {\em BeppoSAX} is around 30\% \citep{fio99,com01} and very 
recently in the 
{\em XMM-Newton} Lockman Hole deep pointing about 60\% is reached 
\citep{has01}.\\
The spectroscopic follow up of the objects making the XRB find predominantly 
Active Galactic Nuclei (AGN).
In the soft band, where optical spectroscopy has reached a high degree
of completeness, the predominant fraction is made by unabsorbed AGN
(type-1 Seyferts and QSOs), with a small fraction of absorbed AGN (essentially 
type-2 Seyferts) \citep{bow96,sch98,zam99}.
The fraction of absorbed type-2 AGN rises if we consider the spectroscopic
identifications of hard X-ray sources in {\em BeppoSAX}, {\em ASCA} and 
{\em Chandra} surveys
\citep{laf01,fio01a,aki00,del00,bar01,toz01}, although the optical follow 
up is far from being complete.\\
The X-ray and optical observations are consistent with current 
XRB synthesis models \citep*{set89,com95,gil01},
which explain the hard XRB spectrum with an appropriate mixture of absorbed 
and unabsorbed AGN, by introducing the corresponding luminosity function 
and cosmological evolution.
In this framework, \citet{fab99} infer an absorption-corrected 
black hole mass density consistent with that estimated from direct optical 
and X-ray studies of nearby unobscured AGN. 
This result requires that most of the X-ray luminosity from AGN ($\sim80\%$)
is absorbed by surrounding gas and probably re-emitted in the infrared band.\\
However synthesis models are far from being unique, depending on a large 
number of hidden parameters. They require, in particular, the presence of a 
significant population of heavily obscured powerful quasars (type-2 QSOs).
Type-2 QSOs have been revealed first by {\em ASCA} and {\em BeppoSAX} 
\citep{oht96,vig99,fra00} and are starting to be discovered 
at high redshift by {\em Chandra} \citep{fab00,nor01}.
These objects are rare (so far, only a few type-2 QSOs are known), 
luminous and hard (heavily absorbed in the soft band). A good way of 
finding them is to perform surveys in the hard X-ray bands, covering 
large solid angles.
The large throughput and effective area, particularly in the harder bands, 
make {\em XMM-Newton} currently the best satellite to perform hard X-ray
surveys.\\
In this paper we present an {\em XMM-Newton} medium-deep
survey covering nearly three square degrees, one of its main goals 
is to constrain the contribution of
absorbed AGN to the XRB. We first overview the data 
preparation (Section~\ref{dataprep})
and the source detection (Section~\ref{srcdet}) procedures, describing then 
the survey characteristics (Section~\ref{sample}) and the first purely X-ray 
results we obtain from the Log(N)-Log(S) (Section~\ref{sectlogn}).
An extensive analysis of the X-ray broad-band properties of the sources and
the optical follow-up of a hard X-ray selected sample will be the subjects
of forthcoming papers (Baldi et al. in prep.)

\section{Data preparation} \label{dataprep}

The survey data are processed using the {\em XMM-Newton} Science Analysis 
System ({\em XMM-SAS}) 
v5.0.\\ 
Before processing, all the datasets have been supplied with the attitude of the 
satellite, which can be considered stable within one arcsecond during any given
observation. Thus, a
good calibration of the absolute celestial positions (within $2^{\prime\prime}
-3^{\prime\prime}$) 
has been obtained from
the pointing coordinates in the Attitude History Files (AHF).
Standard {\em XMM-SAS} tasks $epproc$ and $emproc$ are used to linearize the 
{\em pn} and 
{\em MOS} camera event files.\\
The event files are cleaned up from two further effects,
hot pixels and soft proton flares, both worsening 
data quality.\\ 
The hot and flickering pixel and the 
bad column phenomena, partly due to the electronics of the
detectors, consist basically in the non-X-ray
switching-on of some pixels during an observation and may cause spurious
source detections. 
The majority of them are removed by the {\em XMM-SAS}; we localize the
remaining using the IRAF\footnote{IRAF is distributed
by KPNO, NOAO, operated by the AURA, Inc., for the National Science Foundation.}
task $cosmicrays$ and remove
all the events matching their positions using the multipurpose {\em XMM-SAS} 
task $evselect$.\\ 
Soft proton flares are due to protons with energies less than a 
few hundred keV hitting the detector surface. These particles strongly
enhance the background during an observation; for example $\sim40\%$ of
the long Lockman Hole observation was affected by them. The background 
enhancement forces us to 
completely reject these time intervals with the net effect of a substantial 
reduction of the good integration time.
We locate flares analyzing the light curves at energies higher than 10 keV (in
order to avoid contribution from real X-ray source variability), setting
a threshold for good time intervals at 0.15 cts/s for each {\em MOS} unit and 
at 0.35 cts/s for the {\em pn} unit.

\section{Source detection} \label{srcdet}

The clean linearized event files are used to generate {\em MOS1}, {\em MOS2} 
and {\em pn} images in four different bands: 
0.5-2 keV, 2-10 keV, 2-4.5 keV and 4.5-10 keV.
All the images are built up with a spatial binning of 4.35 arcseconds per
pixel, roughly matching the physical binning of
the {\em pn} images ($4^{\prime\prime}$ pixels) and a factor of about four larger than that of 
the {\em MOS} images ($1.1^{\prime\prime}$ pixels). In any case, the image binning
does not worsen {\em XMM-Newton} spatial resolution, which depends almost
exclusively from the point spread function (PSF).\\
A corresponding set of exposure maps is generated to account for 
spatial quantum efficiency, mirror vignetting and field of view of each
instrument, running {\em XMM-SAS} task $eexpmap$.
This task evaluates the above quantities 
assuming an event energy which corresponds to the mean of the energy boundaries.
In the 2-10 keV band, which covers a wide range of energies, this
may lead to inaccuracies in the estimate of these key quantities. Thus we 
create the 2-10 keV band exposure map as a weighted mean between the 2-4.5 
keV and the 4.5-10 keV exposure maps, assuming an underlying power-law 
spectral model with photon index 1.7.\\
The excellent relative astrometry between the three cameras 
(within $1^{\prime\prime}$, well under the FWHM of the PSF)
allows us to merge together the {\em MOS} and {\em pn} images in order to 
increase the signal-to-noise ratio of the sources and reach fainter 
X-ray fluxes; the corresponding exposure maps are merged too.\\
The source detection and characterization procedure applied to the image sets 
involves the creation of a background map, for each energy band. The first 
step is to run an {\em XMM-SAS} $eboxdetect$ local
detection (in each band independently) to create a source list. Then 
{\em XMM-SAS} $esplinemap$ removes from the original merged image (within 
a radius of 1.5 times the FWHM of the PSF) all the sources in the list
and creates a background map fitting the remaining (the so-called 
{\em cheesed image})
with a cubic spline. Unfortunately, even using the
maximum number of spline nodes (20), the fit is not sufficiently flexible 
to reproduce the local variations of the background. Thus we correct the 
background map pixel by pixel, measuring the counts in the {\em cheesed image}
($cts_{ch}$) and in the background map itself ($cts_{bk}$),
within three times the radius corresponding to an encircled
energy fraction (EEF) of the PSF of $\alpha=0.68$ (hereafter $r_{0.68}$).
We create a corrected background map by multiplying the original image 
by a correction factor which is the $cts_{ch}$ to $cts_{bk}$ ratio. 
After some tests, the radius of $3 r_{0.68}$ has been considered a good 
compromise between taking too many or too few background fluctuations.\\
A preliminary $eboxdetect$ local mode detection run, performed
simultaneously in each energy band, creates the list of candidate 
sources on which to carry out the characterization procedure.\\
Each candidate source is characterized within a radius $r_{0.68}$,
evaluating the source counts $S$ and error $\sigma_S$ \citep[using the 
formula of][]{geh86} following the formulas:
$$
S=\frac{cts_{img}-cts_{bkg}}{\alpha}\:,\:\:\:\:\:\:\:\:\:\:
\sigma_S=\frac{1+\sqrt{cts_{img}+0.75}}{\alpha}\:,
$$
where $cts_{img}$ are the counts (source + background) within $r_{0.68}$ in 
the image and $cts_{bkg}$ are the background counts in the same area in the 
background map. The count rate is then:
$$
cr=\frac{S}{T_{MOS1}+T_{MOS2}+T_{pn}}\:,
$$
where $T_{MOS1}$, $T_{MOS2}$ and $T_{pn}$ are the exposure times of 
the three instruments computed from the exposure maps.\\
The count rate-to-flux conversion factors are computed for each instrument
using the latest response matrices and
assuming a power-law spectral model with photon index 1.7
and galactic $N_H$. The total conversion factor $cf$ has been calculated
using the exposure times for {\em MOS1}, {\em MOS2} and {\em pn}, the 
conversion factors
for the three instruments, $cf_{MOS1}$, $cf_{MOS2}$ and $cf_{pn}$,
following the formula:
$$
\frac{T_{tot}}{cf}=\frac{T_{MOS1}}{cf_{MOS1}}+\frac{T_{MOS2}}{cf_{MOS2}}+\frac{T_{pn}}{cf_{pn}}\:,
$$
where $T_{tot}=(T_{MOS1}+T_{MOS2}+T_{pn}$). 
The source flux is straightforwardly:
$$
F_x=cf\cdot cr\:.
$$ 
For each source we compute
$p$, the probability that counts originate from a
background fluctuation, using Poisson's formula:
$$
\sum_{n=cts_{img}}^\infty e^{-cts_{bkg}}\frac{cts_{bkg}^n}{n!}>p\:;
$$\\
we choose a threshold of $p=2\times10^{-4}$ to decide whether to accept or
not a detected source.\\

\section{The survey} \label{sample}

\begin{figure}
\plotone{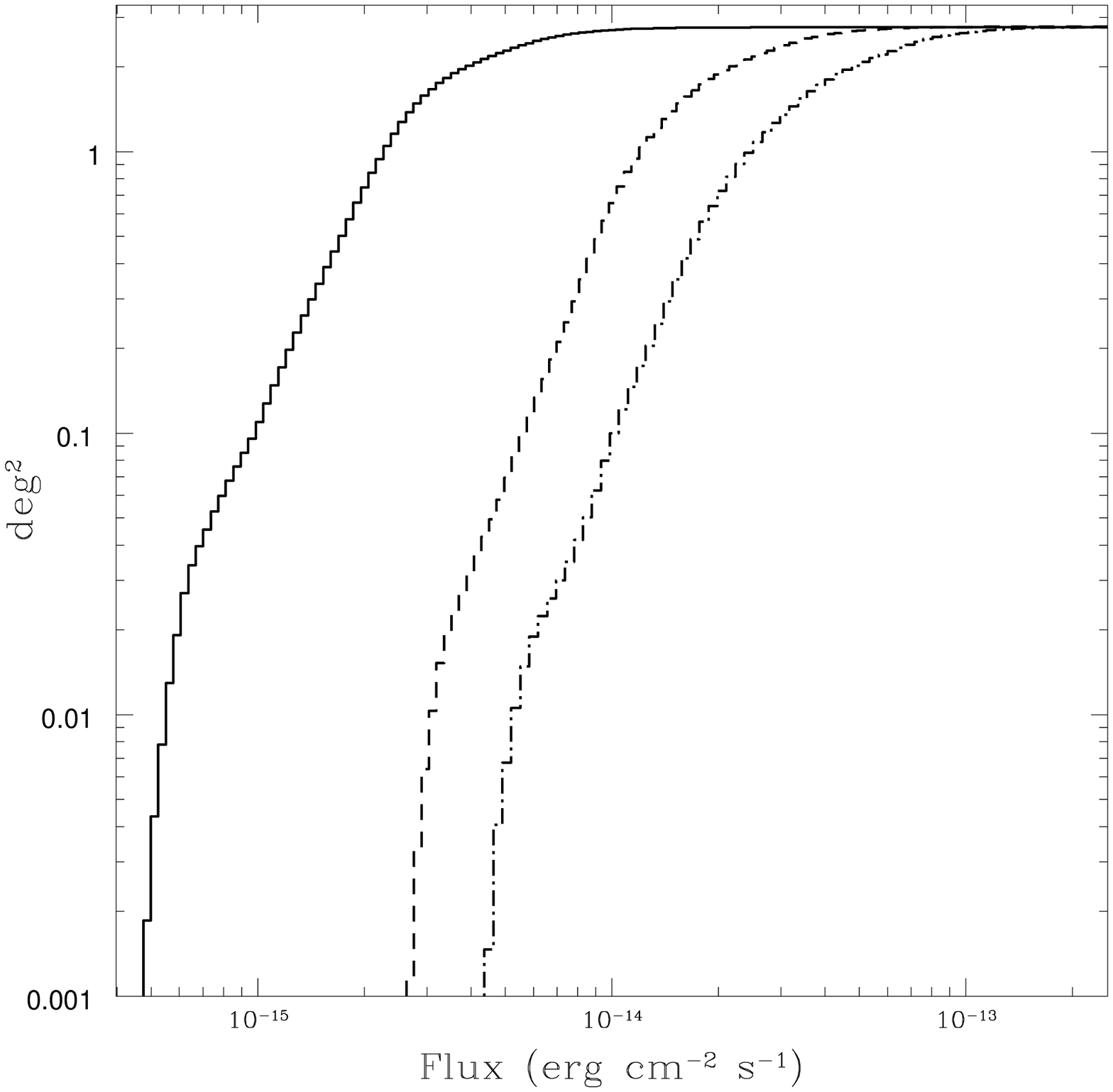}
\caption{The total sky coverage of the survey in the 0.5-2 keV (solid line), 
2-10 keV (dashed line) and 4.5-10 keV band (dot-dashed line). \label{skycov}}
\end{figure}

Our survey covers 15 {\em XMM-Newton} calibration and
performance verification phase fields. The pointings and their characteristics
are listed in Table~\ref{campi}. All fields are at high galactic latitude 
($|bII|$ $>$ 27$^o$), in order to minimize
contamination from galactic sources, have low galactic $N_H$ and at 
least 15 ksec of good integration time.\\
The sky coverage of the sample has been computed using the exposure
maps of each instrument, the background map of the merged image and a model for 
the PSF. We adopt the off-axis angle dependent
PSF model implemented in {\em XMM-SAS} $eboxdetect$ task.\\ 
At each 
image pixel $(x,y)$ we evaluate,
within a radius $r_{0.68}$, the total background counts (from
the background map). From these we calculate 
the minimum total counts (source + background) necessary for a source 
to be detected at a probability $p=2\times10^{-4}$ (defined in 
Section~\ref{srcdet}).
The mean exposure times for {\em MOS1}, {\em MOS2} and {\em pn}, evaluated 
from the exposure maps within $r_{0.68}$, are used to compute the count rate 
$cr$. From the
count rate-to-flux conversion factor $cf$ (computed as in 
Section~\ref{srcdet}) we build a flux limit map and straightforwardly
calculate the sky coverage of a single field.\\
Summing the contribution from all fields
we obtain the total sky coverage of the survey, which is plotted in 
Figure~\ref{skycov}, in three different energy bands.

\section{Log(N)-Log(S)} \label{sectlogn}

\begin{figure}
\epsscale{0.90}
\plotone{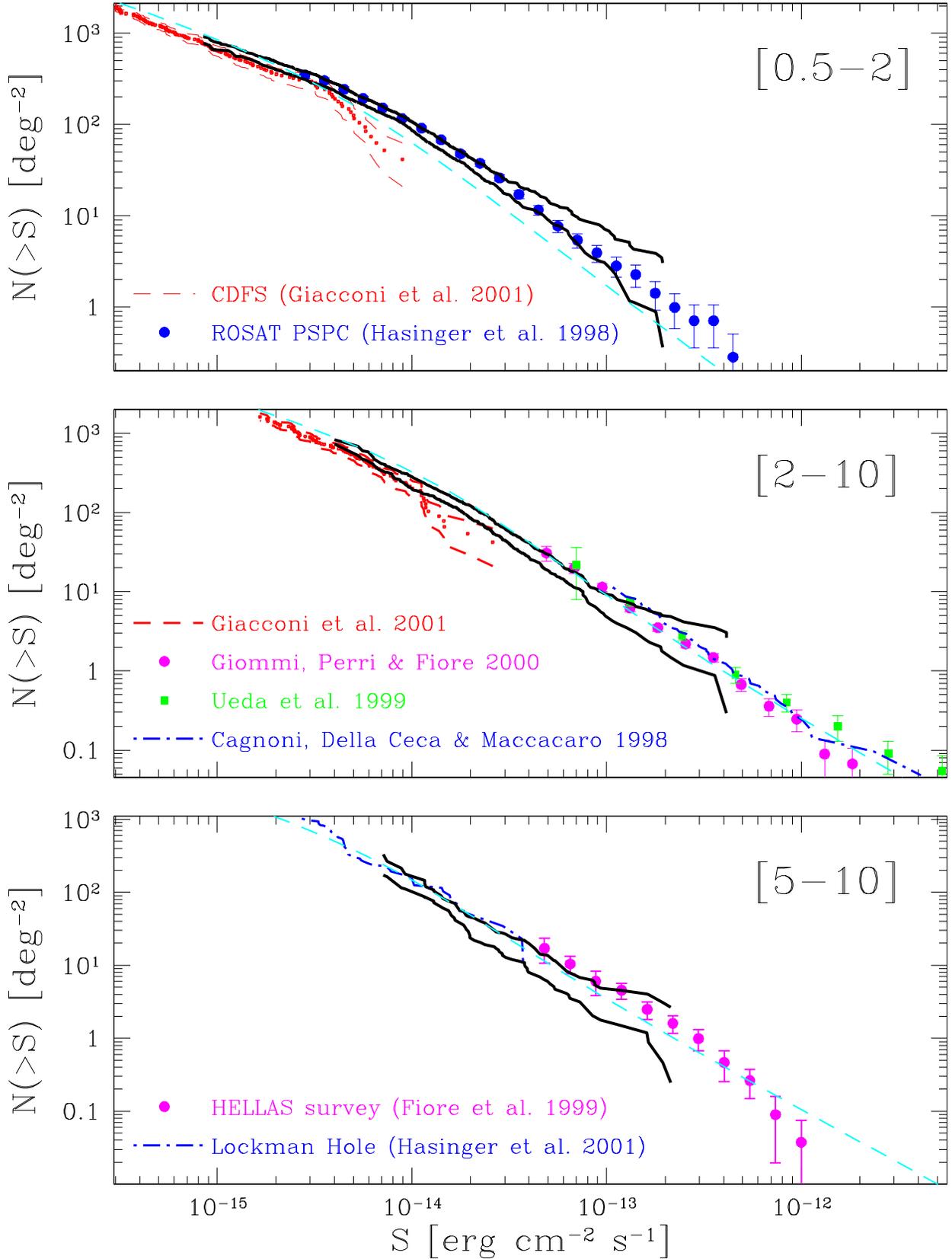}
\caption{The cumulative Log(N)-Log(S) in the 0.5-2 keV (top), 
2-10 keV (center) and 5-10 keV band (bottom). In all diagrams
the black thick solid lines are the upper and lower limits of our Log(N)-Log(S),
computed taking into account systematic effects, as described in 
Section~\ref{sectlogn}. The cyan dashed line represents the predictions
of the improved \citet{com95} XRB synthesis model (see details in text). 
\label{logn}}
\end{figure}

\begin{figure}
\plotone{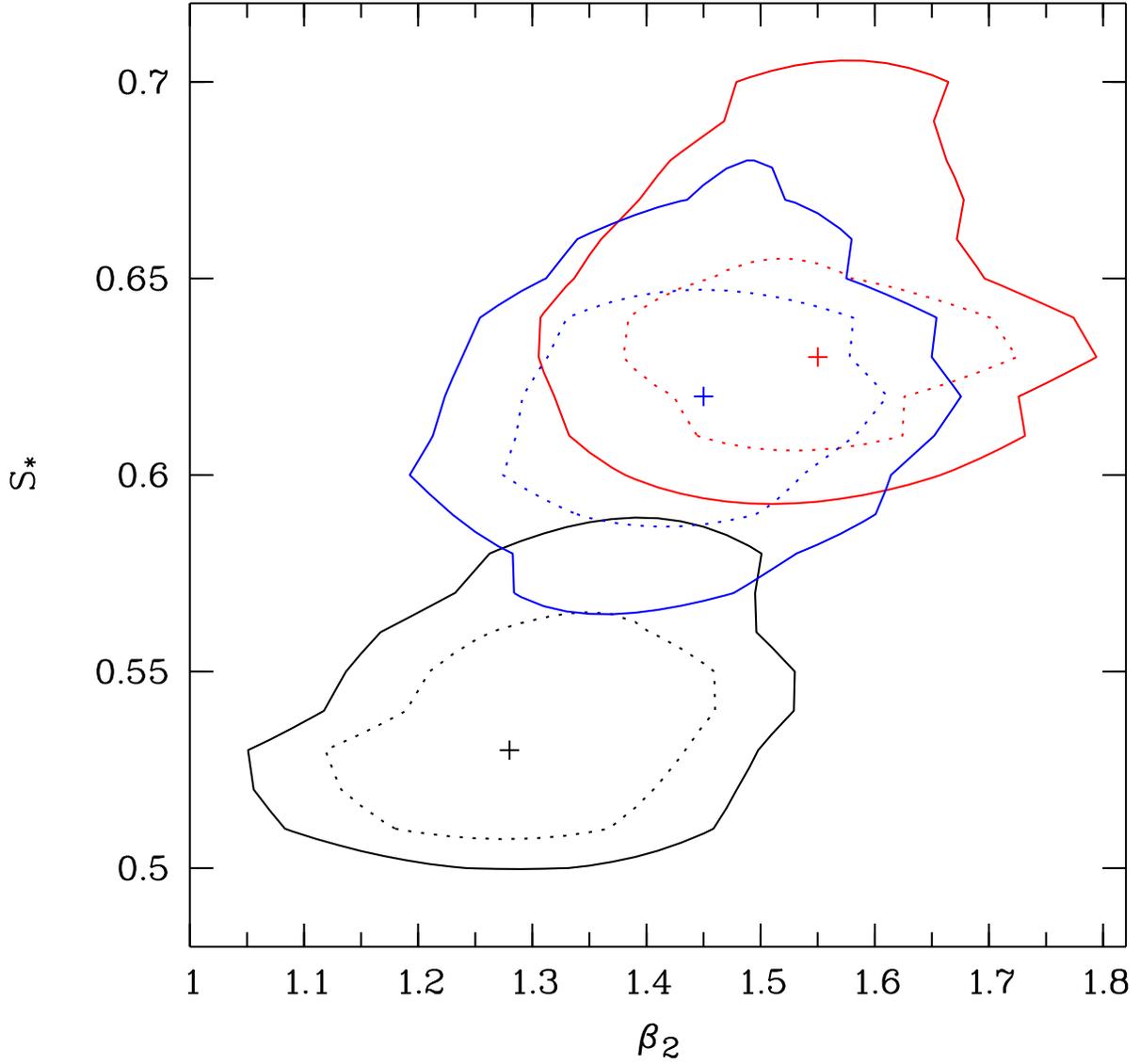}
\caption{Maximum likelihood fit parameters $\beta_2$ and $S_\ast$ (in units of
$10^{-14}$ $erg$ $cm^{-2}$ $s^{-1}$) to the 0.5-2 keV Log(N)-Log(S) for a broken 
power-law model (see text). 
Confidence contours are at 68\% (dashed line) and at 90\% (solid line) for
two interesting parameters. The black contours are computed using a source 
characterization radius corresponding to an EEF of $\alpha=0.68$ and a source
detection probability level of $p=2\times10^{-4}$ for the Log(N)-Log(S). 
The red contours refer to $\alpha=0.80$ and $p=2\times10^{-4}$ while the 
blue contours are computed with $\alpha=0.68$ and $p=2\times10^{-5}$. 
\label{confid}}
\end{figure}

The cumulative Log(N)-Log(S) distribution for our survey has been computed 
by summing up the contribution of each source, weighted by the area in which 
the source could have been detected, following the formula:
$$
N(>S)=\sum_{S_i>S} \frac{1}{\Omega_i},
$$
where $N(>S)$ is the surface number density of sources with flux larger than 
$S$, $S_i$ is the flux of the $i$th source and $\Omega_i$ is the associated 
solid angle.\\
It is worth noting that {\em XMM-Newton} calibrations are not yet fully stable
and systematic errors in the determination of the Log(N)-Log(S) could arise, for
instance, from inaccuracies in the determination of the PSF. Moreover, 
non-poissonian background fluctuations,
at the probability level we have chosen, may cause spurious source detection,
introducing further uncertainties. To account for these effects, we have
computed the Log(N)-Log(S) also using a radius corresponding to an EEF of the 
PSF of 0.80 (instead of 0.68) for the source characterization and a more 
stringent probability threshold of $p=2\times10^{-5}$ (instead of 
$p=2\times10^{-4}$). The different curves
we obtain (and relative 1$\sigma$ statistical uncertainties) are used to 
determine the upper and lower limits of the Log(N)-Log(S), plotted in 
Figure~\ref{logn}. The Log(N)-Log(S) distributions contain
1022 sources, 495 sources and 100 sources, for the 0.5-2 keV, 2-10
keV and 5-10 keV band, respectively (using $p=2\times10^{-4}$ and EEF=0.68). 
It is worth noting that we compute the
Log(N)-Log(S) in the 5-10 keV instead of the 4.5-10 keV band for consistency
with previous works \citep{fio01b,has01}. We correct the 4.5-10 keV fluxes
to obtain the 5-10 keV fluxes, assuming an underlying power-law spectral 
model with galactic $N_H$ and photon index $\Gamma=1.7$.\\
In the soft band (0.5-2 keV), where we have one of the largest samples to 
date, the data are in 
agreement, within the errors, with both {\em ROSAT PSPC} Lockman Hole data 
\citep{has98} and {\em Chandra} Deep Field 
South data \citep[CDFS; ][]{gia01}. In this band we go about a factor of
four deeper than {\em ROSAT PSPC} data, although obviously not as deep as 
{\em Chandra} in the CDFS. The Log(N)-Log(S) shows a clear flattening 
starting from fluxes around $10^{-14}$ erg cm$^{-2}$ s$^{-1}$. A similar 
behaviour has been already observed in {\em ROSAT} data \citep{has98}.
A possible 
explanation for it may reside in the luminosity dependent density 
evolution (LDDE) models of the soft X-ray AGN luminosity function developed 
on {\em ROSAT} data by \citet*{miy00}.\\
We fit the soft Log(N)-Log(S) distribution 
with a single power-law model in the form $N(>S)=KS_{14}^{-\alpha}$ ($S_{14}$
is the flux in units of $10^{-14}$ erg cm$^{-2}$ s$^{-1}$), using a 
maximum likelihood method \citep*{cra70,mur73}. 
This method has the advantage of using directly the unbinned
data. The likelihood has a maximum at a slope  
$\alpha=0.93\pm0.05$ and the corresponding normalization of the curve is 
$K=80.8^{+6.4}_{-5.2}$ (the errors have been computed not only considering
statistical uncertainties but also the scatter between the three different
Log(N)-Log(S) described earlier in the text).
However a single power-law model can be rejected applying a K-S
test which gives a probability $<10^{-3}$. We consider then a broken
power-law model for the differential Log(N)-Log(S), defined as
$$
\frac{dN}{dS}=\left\{
\begin{array}{cc}
 k_1S_{14}^{-\beta_1} &\:  S>S_\ast\\
 k_2S_{14}^{-\beta_2} &\:  S<S_\ast\\
\end{array}
\right.
$$
where $\beta_1$ is the power-law index at brighter fluxes, 
$\beta_2$ the index at fainter fluxes, $S_\ast$ is the flux of the 
break, $k_1$ and $k_2$ are the normalization factors 
($k_2=k_1S_\ast^{\beta_2-\beta_1}$ to have continuity in the differential 
counts).
Applying the maximum likelihood fit to the data, we obtain a best-fit 
value of $\beta_1=2.21^{+0.06}_{-0.09}$ (with a corresponding normalization
$k_1=118.8^{+13.9}_{-11.1}$), 
while the confidence contours for $\beta_2$ and $S_\ast$, for each of the three
Log(N)-Log(S) curves described earlier in this Section, are plotted in 
Figure~\ref{confid}.
The break flux $S_\ast$, at 1$\sigma$ confidence level for two interesting
parameters, ranges in a narrow 
interval of values, between $5\times10^{-15}$ and $6.5\times10^{-15}$ 
erg cm$^{-2}$ s$^{-1}$. The differential slope at fainter fluxes $\beta_2$
is not tightly constrained, ranging between 1.1 and 1.7 (1$\sigma$
confidence level for two interesting parameters). In any case, these values 
of $\beta_2$ are somewhat lower
than those found by \citet{has98} fitting the {\em ROSAT} data. 
The above authors find also a break
at brighter fluxes: the discrepancy 
could arise from the fact that we are observing a fainter and flatter part 
of the Log(N)-Log(S), which was not accessible with the ROSAT PSPC data.\\
In the 2-10 keV energy band, we certainly have the largest hard X-ray selected
sample available to date at these fluxes.
Also in this case the data are in good agreement with
previous determinations,
by {\em BeppoSAX} \citep{gio00} and {\em ASCA} \citep{cag98,ued99},
in the brighter part, and by {\em Chandra}
\citep{gia01} in the fainter part. In this band, our Log(N)-Log(S)
nicely fills in the gap between the {\em Chandra} deep surveys and
the shallow {\em BeppoSAX} and {\em ASCA} surveys. A slight slope flattening
(around $2\times10^{-14}$ erg cm$^{-2}$ s$^{-1}$) comes out also in the 2-10
keV Log(N)-Log(S). A similar flattening has already been observed by
\citet{has01} in
the Lockman Hole {\em XMM-Newton} deep observations.
A maximum
likelihood fitting technique has been applied also to the 2-10 keV 
Log(N)-Log(S). A single power-law model has its best-fit value at
$\alpha=1.34^{+0.11}_{-0.10}$ and a normalization $K=229.2^{+29.3}_{-19.6}$. 
The K-S probability ($>10\%$)
do not allow us to reject the model indicating that the flattening is not 
particularly significant. However the best-fit value of the slope is 
significantly sub-euclidean, in contrast to BeppoSAX and ASCA findings, 
indicating that probably the Log(N)-Log(S) flattens at faint fluxes.\\
The 5-10 keV Log(N)-Log(S) is in agreement,
within the errors, with both {\em XMM-Newton} Lockman Hole data
\citep{has01}, which is a subsample of ours, and {\em BeppoSAX} HELLAS
survey \citep{fio01b}. Our Log(N)-Log(S) connects {\em XMM-Newton} deep
observations with shallower {\em BeppoSAX} ones. The sample selected in this
band (100 sources) is currently smaller than the {\em BeppoSAX} HELLAS sample
(about 150 sources). However, we go deeper by an order of magnitude than the
HELLAS survey and the error circle we can use in the optical
follow up (conservatively we are assuming $3^{\prime\prime}$) is considerably 
smaller than
{\em BeppoSAX} (about $1^\prime$), making the optical identification far easier.\\
A maximum likelihood fit of the 5-10 keV Log(N)-Log(S)
with a single power-law model gives a value of
$\alpha=1.54^{+0.25}_{-0.19}$ and a normalization $K=175.2^{+56.3}_{-36.2}$. 
As in the 2-10 keV band, the single 
power-law model is found to give an acceptable description of the data 
(the K-S probability is larger than $20\%$).\\
In each panel of 
Figure~\ref{logn}, the cyan dashed line represents the expected 
Log(N)-Log(S) from the improved \citet{com95} XRB synthesis model 
\citep[see][for details]{com01}. In the 0.5-2 keV band, the counts overestimates
the model predictions at bright fluxes, because of the 
contribution from clusters and stars to the soft Log(N)-Log(S) .
At fainter fluxes, where the AGN are the dominant contributors, 
the agreement is quite good.
In the 2-10 keV band the agreement between XRB model predictions and our 
Log(N)-Log(S)
is good at brighter fluxes, becoming marginal towards fainter fluxes.
However, by varying the normalization of the model of $\sim20\%$,
the predicted Log(N)-Log(S) agrees well with both our data and CDFS data.
In the 5-10 keV band the model predictions are in agreement within the errors 
with our Log(N)-Log(S) and the Lockman Hole and HELLAS surveys.
\\
It is worth noting that we do not make any correction for confusion or 
Eddington biases. Nevertheless, the agreement between our source counts and
{\em Chandra} and {\em ROSAT} data, in the 0.5-2 keV band, indicates that 
source confusion is still negligible at these fluxes.

\section{Summary}

We have carried out a serendipitous {\em XMM-Newton} survey. We cover 
nearly three square degrees in 15 fields 
observed during satellite
calibration and performance verification phase. This is, 
to date,
the {\em XMM-Newton} survey with the largest solid angle.\\
The present sample is one of the largest available in the 0.5-2 keV
band and is surely the largest in the 2-10 keV band at these fluxes.
In the 4.5-10 keV band we currently have a smaller sample than the {\em BeppoSAX} 
HELLAS survey. However, the flux limit is a factor about 10 deeper than 
HELLAS and the optical follow up of our survey is easier 
because of XMM-Newton better positional accuracy.\\
We computed the Log(N)-Log(S) curves in the 0.5-2 keV, 
2-10 keV and 5-10 keV bands.
Our measurements are in agreement with previous
determinations by other satellites and {\em XMM-Newton} 
itself \citep{has98,ued99,cag98,gio00,gia01,has01} and with the 
predictions of the improved \citet{com95} XRB synthesis model.\\ 
In the hard bands, we 
sample an intermediate flux range: deeper than {\em ASCA} and {\em BeppoSAX} 
and shallower 
than {\em Chandra} and {\em XMM-Newton} deep pencil-beam surveys. It is worth 
to note
that our approach is complementary to the latters: we probe large areas,
at fluxes bright enough to allow, at least, a coarse spectral
characterization of them. One of our main goals is in fact to find 
a good number of those rare objects (like type-2 QSOs) which are supposed to
contribute significantly to the extragalactic hard X-ray background.\\
In the soft band, the Log(N)-Log(S) distribution shows a flattening 
around $5\times10^{-15}$ erg cm$^{-2}$ s$^{-1}$. A similar result was also found
from the {\em ROSAT} data \citep{has98}. A broken power-law fit gives
a differential slope index $\beta_2$ for the fainter part, flatter than
\citet{has98}. The difference probably
results from the fact that we are sampling different parts of the 
Log(N)-Log(S).
A slight slope flattening of the Log(N)-Log(S) is also observed in the 
2-10 keV band, around fluxes of $2\times10^{-14}$ erg cm$^{-2}$ s$^{-1}$, 
although the data are consistent with a single power-law with a cumulative 
slope index 
$\alpha=1.34^{+0.11}_{-0.10}$. A single power-law fit is tenable also
for the 5-10 keV Log(N)-Log(S) and gives a slope 
$\alpha=1.54^{+0.25}_{-0.19}$.\\
An extensive analysis of the X-ray broad-band properties of the sources and
the optical follow-up of a hard X-ray selected sample will be the subjects
of forthcoming papers (Baldi et al. in prep.).

\acknowledgments
We thank A. De Luca for developing the hot pixel cleaning algorithm. 
We are also grateful to G. Zamorani, G. C. Perola and all members of the 
{\sc HELLAS2XMM} team for useful discussions. We also thank the 
referee for useful suggestions which improved the presentation of the 
results. AB and SM acknowledge partial financial support by
ASI I/R/190/00 contract.

\clearpage


\clearpage 

\clearpage 

\clearpage

\begin{deluxetable}{clrrrrr}
\tabletypesize{\small}
\tablecaption{The {\em XMM-Newton} Cal-PV field sample. \label{campi}}
\tablewidth{0pt}
\tablehead{
\colhead{Revs\tablenotemark{a}} &
\colhead{Target} & \colhead{$T_{MOS1}$(ks)\tablenotemark{b}}   & 
\colhead{$T_{MOS2}$(ks)\tablenotemark{c}}   &
\colhead{$T_{pn}$(ks)\tablenotemark{d}} &
\colhead{N$_H$ (cm$^{-2}$)\tablenotemark{e}}  & 
\colhead{$bII$($^o$)\tablenotemark{f}}
}
\startdata
51&PKS0537-286&19.0&37.0&36.6&$2.1\cdot 10^{20}$&-27.3\\
57&PKS0312-770&25.5&25.5&22.1&$8.0\cdot 10^{20}$&-37.6\\
63&MS0737.9+7441&37.3&38.5&31.6&$3.5\cdot 10^{20}$&29.6\\
70-71-73-74-81&Lockman Hole&84.6&86.2&104.9&$5.6\cdot 10^{19}$&53.1\\
75&Mkn 205&29.0&30.6&17.3&$3.0\cdot 10^{20}$&41.7\\
81-88-185&BPM 16274&38.9&39.2&33.0&$3.2\cdot 10^{20}$&-65.0\\
82&MS1229.2+6430&24.6&&24.9&$2.0\cdot 10^{20}$&52.8\\
84-153&PKS0558-504&20.2&20.4&8.4&$4.5\cdot 10^{20}$&-28.6\\
84-165-171&Mkn 421&98.4&116.5&&$7.0\cdot 10^{19}$&65.0\\
88&Abell 2690&17.5&17.5&16.2&$1.9\cdot 10^{20}$&-78.4\\
90&G158-100&21.3&16.6&&$2.5\cdot 10^{20}$&-74.5\\
90&GD153&36.5&21.2&26.2&$2.4\cdot 10^{20}$&84.7\\
97&IRAS13349+2438&41.4&&&$1.2\cdot 10^{20}$&79.3\\
101&Abell 1835&27.7&27.7&22.9&$2.3\cdot 10^{20}$&60.6\\
161&Mkn 509&16.8&16.4&&$4.1\cdot 10^{20}$&-29.9\\
\enddata

\tablenotetext{a}{{\em XMM-Newton} revolution numbers}
\tablenotetext{b}{{\em MOS1} good integration time}
\tablenotetext{c}{{\em MOS2} good integration time}
\tablenotetext{d}{{\em pn} good integration time}
\tablenotetext{e}{Galactic Hydrogen column density \citep{sta92}}
\tablenotetext{f}{Galactic latitude}

\end{deluxetable}

\end{document}